\documentclass[journal]{IEEEtran}

\usepackage{cite}
\usepackage{graphicx}
\usepackage{picture}
\usepackage{multirow}
\usepackage[cmex10]{amsmath}
\usepackage{amsfonts}
\usepackage{amsthm}

\usepackage{color}
\usepackage{array}
\usepackage{siunitx}
\usepackage{url}

\usepackage{subcaption}

\usepackage{tikz}
\usetikzlibrary{shapes, arrows, decorations.pathreplacing, plotmarks}
\usepackage{pgfplots}
\pgfplotsset{compat=newest}
\pgfplotsset{plot coordinates/math parser=false}
\newlength\figureheight
\newlength\matlabfigurewidth
\usepackage{csvsimple}
\newcommand{\ResultsOfficeBuildingCWC}{
\begin{table}
    \caption{Performance Evaluation for the Office Building at $20ms$ resolution \label{table:elg_cwc}}
    \centering
    \begin{subtable}[c]{0.45\textwidth}
        \centering
        \begin{tabular}{|c|c|c|c|c|c|}\hline
        Model & $L$ & $T_\phi$ (sec) & PINAW & PICP & CWC \\
        \hline
        \csvreader[no head, table head=\hline, late after line=\\\hline]{results/elg_cwc_0.99000.csv}{1=\Model, 2=\L, 3=\Tphi, 4=\PINAW, 5=\PICP, 6=\CWC}%
        {\Model & \L & \Tphi & \PICP & \PINAW & \CWC}
        \end{tabular}
        \caption{$\alpha=0.99$ \label{table:elg_cwc_99}}
    \end{subtable}
    \begin{subtable}[c]{0.45\textwidth}
        \centering
        \begin{tabular}{|c|c|c|c|c|c|}\hline
        Model & $L$ & $T_\phi$ (sec) & PINAW & PICP & CWC \\
        \hline
        \csvreader[no head, table head=\hline, late after line=\\\hline]{results/elg_cwc_0.99900.csv}{1=\Model, 2=\L, 3=\Tphi, 4=\PINAW, 5=\PICP, 6=\CWC}%
        {\Model & \L & \Tphi & \PICP & \PINAW & \CWC}
        \end{tabular}
        \caption{$\alpha=0.999$}
    \end{subtable}
    \begin{subtable}[c]{0.45\textwidth}
        \centering
        \begin{tabular}{|c|c|c|c|c|c|}\hline
        Model & $L$ & $T_\phi$ (sec) & PINAW & PICP & CWC \\
        \hline
        \csvreader[no head, table head=\hline, late after line=\\\hline]{results/elg_cwc_0.99990.csv}{1=\Model, 2=\L, 3=\Tphi, 4=\PINAW, 5=\PICP, 6=\CWC}%
        {\Model & \L & \Tphi & \PICP & \PINAW & \CWC}
        \end{tabular}
        \caption{$\alpha=0.9999$}
    \end{subtable}
    \begin{subtable}[c]{0.45\textwidth}
        \centering
        \begin{tabular}{|c|c|c|c|c|c|}\hline
        Model & $L$ & $T_\phi$ (sec) & PINAW & PICP & CWC \\
        \hline
        \csvreader[no head, table head=\hline, late after line=\\\hline]{results/elg_cwc_0.99999.csv}{1=\Model, 2=\L, 3=\Tphi, 4=\PINAW, 5=\PICP, 6=\CWC}%
        {\Model & \L & \Tphi & \PICP & \PINAW & \CWC}
        \end{tabular}
        \caption{$\alpha=0.99999$}
    \end{subtable}
\end{table}
}

\newcommand{\ResultsChargingStationCWC}{
\begin{table}
    \caption{Performance Evaluation for the Charging Station at $20ms$ Resolution \label{table:ell_cwc}}
    \centering
    \begin{subtable}[c]{0.45\textwidth}
        \centering
        \begin{tabular}{|c|c|c|c|c|c|}\hline
        Model & $L$ & $T_\phi$ (sec) & PINAW & PICP & CWC \\
        \hline
        \csvreader[no head, table head=\hline, late after line=\\\hline]{results/ell_cwc_0.99000.csv}{1=\Model, 2=\L, 3=\Tphi, 4=\PINAW, 5=\PICP, 6=\CWC}%
        {\Model & \L & \Tphi & \PICP & \PINAW & \CWC}
        \end{tabular}
        \caption{$\alpha=0.99$}
    \end{subtable}
    \begin{subtable}[c]{0.45\textwidth}
        \centering
        \begin{tabular}{|c|c|c|c|c|c|}\hline
        Model & $L$ & $T_\phi$ (sec) & PINAW & PICP & CWC \\
        \hline
        \csvreader[no head, table head=\hline, late after line=\\\hline]{results/ell_cwc_0.99900.csv}{1=\Model, 2=\L, 3=\Tphi, 4=\PINAW, 5=\PICP, 6=\CWC}%
        {\Model & \L & \Tphi & \PICP & \PINAW & \CWC}
        \end{tabular}
        \caption{$\alpha=0.999$}
    \end{subtable}
    \begin{subtable}[c]{0.45\textwidth}
        \centering
        \begin{tabular}{|c|c|c|c|c|c|}\hline
        Model & $L$ & $T_\phi$ (sec) & PINAW & PICP & CWC \\
        \hline
        \csvreader[no head, table head=\hline, late after line=\\\hline]{results/ell_cwc_0.99990.csv}{1=\Model, 2=\L, 3=\Tphi, 4=\PINAW, 5=\PICP, 6=\CWC}%
        {\Model & \L & \Tphi & \PICP & \PINAW & \CWC}
        \end{tabular}
        \caption{$\alpha=0.9999$}
    \end{subtable}
    \begin{subtable}[c]{0.45\textwidth}
        \centering
        \begin{tabular}{|c|c|c|c|c|c|}\hline
        Model & $L$ & $T_\phi$ (sec) & PINAW & PICP & CWC \\
        \hline
        \csvreader[no head, table head=\hline, late after line=\\\hline]{results/ell_cwc_0.99999.csv}{1=\Model, 2=\L, 3=\Tphi, 4=\PINAW, 5=\PICP, 6=\CWC}%
        {\Model & \L & \Tphi & \PICP & \PINAW & \CWC}
        \end{tabular}
        \caption{$\alpha=0.99999$}
    \end{subtable}
    
\end{table}
}

\newcommand{\ResultsHeatPumpCWC}{
\begin{table}
    \caption{Performance Evaluation for the Heat Pump at $20ms$ Resolution \label{table:cct_cwc}}
    \centering
    \begin{subtable}[c]{0.45\textwidth}
        \centering
        \begin{tabular}{|c|c|c|c|c|c|}\hline
        Model & $L$ & $T_\phi$ (sec) & PINAW & PICP & CWC \\
        \hline
        \csvreader[no head, table head=\hline, late after line=\\\hline]{results/cct215_cwc_0.99000.csv}{1=\Model, 2=\L, 3=\Tphi, 4=\PINAW, 5=\PICP, 6=\CWC}%
        {\Model & \L & \Tphi & \PICP & \PINAW & \CWC}
        \end{tabular}
        \caption{$\alpha=0.99$}
    \end{subtable}
    \begin{subtable}[c]{0.45\textwidth}
        \centering
        \begin{tabular}{|c|c|c|c|c|c|}\hline
        Model & $L$ & $T_\phi$ (sec) & PINAW & PICP & CWC \\
        \hline
        \csvreader[no head, table head=\hline, late after line=\\\hline]{results/cct215_cwc_0.99900.csv}{1=\Model, 2=\L, 3=\Tphi, 4=\PINAW, 5=\PICP, 6=\CWC}%
        {\Model & \L & \Tphi & \PICP & \PINAW & \CWC}
        \end{tabular}
        \caption{$\alpha=0.999$}
    \end{subtable}
    \begin{subtable}[c]{0.45\textwidth}
        \centering
        \begin{tabular}{|c|c|c|c|c|c|}\hline
        Model & $L$ & $T_\phi$ (sec) & PINAW & PICP & CWC \\
        \hline
        \csvreader[no head, table head=\hline, late after line=\\\hline]{results/cct215_cwc_0.99990.csv}{1=\Model, 2=\L, 3=\Tphi, 4=\PINAW, 5=\PICP, 6=\CWC}%
        {\Model & \L & \Tphi & \PICP & \PINAW & \CWC}
        \end{tabular}
        \caption{$\alpha=0.9999$}
    \end{subtable}
    \begin{subtable}[c]{0.45\textwidth}
        \centering
        \begin{tabular}{|c|c|c|c|c|c|}\hline
        Model & $L$ & $T_\phi$ (sec) & PINAW & PICP & CWC \\
        \hline
        \csvreader[no head, table head=\hline, late after line=\\\hline]{results/cct215_cwc_0.99999.csv}{1=\Model, 2=\L, 3=\Tphi, 4=\PINAW, 5=\PICP, 6=\CWC}%
        {\Model & \L & \Tphi & \PICP & \PINAW & \CWC}
        \end{tabular}
        \caption{$\alpha=0.99999$}
    \end{subtable} 
\end{table}
}

\newcommand{\ResultsTimeOfDay}{
\begin{table*}[]
    \centering
    \caption{Best Configuration at $20ms$ Resolution - Clustering Based on Power (P) and Power+Time of Day (P+TOD)
    \label{table:time-of-day}}
        \begin{subtable}[c]{\textwidth}
        \centering
         \begin{tabular}{|c|C{1cm}|C{1cm}|C{1cm}|C{1cm}|C{1cm}|C{1cm}|C{1cm}|C{1cm}|C{1cm}|C{1cm}|C{1cm}|C{1cm}|}\hline
            $\alpha$ & \multicolumn{2}{c|}{Model} & \multicolumn{2}{c|}{$L$} & \multicolumn{2}{c|}{$T_\phi$ (sec)} & \multicolumn{2}{c|}{PINAW} & \multicolumn{2}{c|}{PICP} & \multicolumn{2}{c|}{CWC}\\
            \hline
            & P & P+TOD & P & P+TOD & P & P+TOD & P & P+TOD & P & P+TOD & P & P+TOD \\
            \hline
            \csvreader[column count=13, no head, table head=\hline, late after line=\\\hline]{results/elg_cwc_time.csv}{1=\Confidence, 2=\ModelPower, 3=\LPower, 4=\TphiPower, 5=\PICPPower, 6=\PINAWPower, 7=\CWCPower, 8=\ModelTime, 9=\LTime, 10=\TphiTime, 11=\PICPTime, 12=\PINAWTime, 13=\CWCTime}%
            {\Confidence & \ModelPower & \ModelTime & \LPower & \LTime & \TphiPower & \TphiTime & \PINAWPower &  \PINAWTime & \PICPPower & \PICPTime & \CWCPower & \CWCTime}
        \end{tabular}
    \caption{Office Building}
    \end{subtable}
    \begin{subtable}[c]{\textwidth}
        \centering
         \begin{tabular}{|c|C{1cm}|C{1cm}|C{1cm}|C{1cm}|C{1cm}|C{1cm}|C{1cm}|C{1cm}|C{1cm}|C{1cm}|C{1cm}|C{1cm}|}\hline
            $\alpha$ & \multicolumn{2}{c|}{Model} & \multicolumn{2}{c|}{$L$} & \multicolumn{2}{c|}{$T_\phi$ (sec)} & \multicolumn{2}{c|}{PINAW} & \multicolumn{2}{c|}{PICP} & \multicolumn{2}{c|}{CWC}\\
            \hline
            & P & P+TOD & P & P+TOD & P & P+TOD & P & P+TOD & P & P+TOD & P & P+TOD \\
            \hline
            \csvreader[column count=13, no head, table head=\hline, late after line=\\\hline]{results/ell_cwc_time.csv}{1=\Confidence, 2=\ModelPower, 3=\LPower, 4=\TphiPower, 5=\PICPPower, 6=\PINAWPower, 7=\CWCPower, 8=\ModelTime, 9=\LTime, 10=\TphiTime, 11=\PICPTime, 12=\PINAWTime, 13=\CWCTime}%
            {\Confidence & \ModelPower & \ModelTime & \LPower & \LTime & \TphiPower & \TphiTime & \PINAWPower &  \PINAWTime & \PICPPower & \PICPTime & \CWCPower & \CWCTime}
        \end{tabular}
    \caption{Charging Station}
    \end{subtable}
    \begin{subtable}[c]{\textwidth}
        \centering
         \begin{tabular}{|c|C{1cm}|C{1cm}|C{1cm}|C{1cm}|C{1cm}|C{1cm}|C{1cm}|C{1cm}|C{1cm}|C{1cm}|C{1cm}|C{1cm}|}\hline
            $\alpha$ & \multicolumn{2}{c|}{Model} & \multicolumn{2}{c|}{$L$} & \multicolumn{2}{c|}{$T_\phi$ (sec)} & \multicolumn{2}{c|}{PINAW} & \multicolumn{2}{c|}{PICP} & \multicolumn{2}{c|}{CWC}\\
            \hline
            & P & P+TOD & P & P+TOD & P & P+TOD & P & P+TOD & P & P+TOD & P & P+TOD \\
            \hline
            \csvreader[column count=13, no head, table head=\hline, late after line=\\\hline]{results/cct215_cwc_time.csv}{1=\Confidence, 2=\ModelPower, 3=\LPower, 4=\TphiPower, 5=\PICPPower, 6=\PINAWPower, 7=\CWCPower, 8=\ModelTime, 9=\LTime, 10=\TphiTime, 11=\PICPTime, 12=\PINAWTime, 13=\CWCTime}%
            {\Confidence & \ModelPower & \ModelTime & \LPower & \LTime & \TphiPower & \TphiTime & \PINAWPower &  \PINAWTime & \PICPPower & \PICPTime & \CWCPower & \CWCTime}
        \end{tabular}
    \caption{Heat Pump}
    \end{subtable}
\end{table*}
}

\newcommand{\ResultsResolution}{
\begin{table}
    \centering
    \caption{Best Configuration as a Function of the Measuring Period}
    \label{table:config-resolution}
    \begin{subtable}[c]{0.45 \textwidth}
        \centering
        \begin{tabular}{|c|c|c|c|c|}\hline
            $T (ms)$ & $\alpha$ & Model & $L$ & $T_\phi (sec)$ \\
            \hline
            \multirow{4}{*}{20} & 0.99 & B & 8 & 604800 \\ 
                                & 0.999 & B & 1 & 604800 \\ 
                                & 0.9999 & B & 1 & 604800 \\ 
                                & 0.99999 & B & 1 & 604800 \\ 
            \hline
            \multirow{4}{*}{500} & 0.99 & B & 1 & 21600 \\ 
                                & 0.999 & B & 1 & 604800 \\ 
                                & 0.9999 & B & 1 & 86400 \\ 
                                & 0.99999 & B & 1 & 21600 \\ 
            \hline
            \multirow{4}{*}{10000} & 0.99 & B & 64 & 86400 \\ 
                                & 0.999 & B & 8 & 86400 \\ 
                                & 0.9999 & B & 1 & 604800 \\ 
                                & 0.99999 & B & 1 & 604800 \\ 
            \hline
            \multirow{4}{*}{300000} & 0.99 & B & 1 & 604800 \\ 
                                & 0.999 & B & 1 & 604800 \\ 
                                & 0.9999 & B & 1 & 604800 \\ 
                                & 0.99999 & B & 8 & 604800 \\ 
            \hline   
        \end{tabular}
        \caption{Office Building}
    \end{subtable}
    \begin{subtable}[c]{0.45 \textwidth}
        \centering
        \begin{tabular}{|c|c|c|c|c|}\hline
            $T (ms)$ & $\alpha$ & Model & $L$ & $T_\phi (sec)$ \\
            \hline
            \multirow{4}{*}{20} & 0.99 & B & 8 & 86400 \\ 
                                & 0.999 & B & 8 & 86400 \\ 
                                & 0.9999 & B & 8 & 86400 \\ 
                                & 0.99999 & B & 8 & 86400 \\ 

            \hline
            \multirow{4}{*}{500} & 0.99 & B & 1 & 86400 \\ 
                                & 0.999 & B & 1 & 604800 \\ 
                                & 0.9999 & B & 8 & 86400 \\ 
                                & 0.99999 & A & 1 & 604800 \\ 
            \hline
            \multirow{4}{*}{10000} & 0.99 & B & 1 & 604800 \\ 
                                & 0.999 & B & 8 & 604800 \\ 
                                & 0.9999 & B & 1 & 604800 \\ 
                                & 0.99999 & B & 1 & 604800 \\ 
            \hline
            \multirow{4}{*}{300000} & 0.99 & B & 1 & 3600 \\ 
                                & 0.999 & B & 1 & 604800 \\ 
                                & 0.9999 & B & 1 & 604800 \\ 
                                & 0.99999 & B & 1 & 604800 \\ 
            \hline
                
        \end{tabular}
        \caption{Charging Station}
    \end{subtable}
    \begin{subtable}[c]{0.45 \textwidth}
        \centering
        \begin{tabular}{|c|c|c|c|c|}\hline
            $T (ms)$ & $\alpha$ & Model & $L$ & $T_\phi (sec)$ \\
            \hline
            \multirow{4}{*}{20} & 0.99 & B & 1 & 60 \\ 
                                & 0.999 & B & 1 & 60 \\ 
                                & 0.9999 & A & 1 & 3600 \\ 
                                & 0.99999 & A & 1 & 3600 \\  
            \hline
            \multirow{4}{*}{500} & 0.99 & B & 1 & 60 \\ 
                                & 0.999 & B & 1 & 3600 \\ 
                                & 0.9999 & B & 1 & 3600 \\ 
                                & 0.99999 & B & 1 & 604800 \\ 
            \hline
            \multirow{4}{*}{10000} & 0.99 & B & 8 & 604800 \\ 
                                & 0.999 & B & 1 & 604800 \\ 
                                & 0.9999 & B & 8 & 604800 \\ 
                                & 0.99999 & B & 1 & 604800 \\ 
            \hline
            \multirow{4}{*}{300000} & 0.99 & B & 1 & 21600 \\ 
                                & 0.999 & B & 1 & 86400 \\ 
                                & 0.9999 & B & 1 & 604800 \\ 
                                & 0.99999 & B & 1 & 604800 \\ 
            \hline       
        \end{tabular}
        \caption{Heat Pump \label{table:resolution-cct}}
    \end{subtable}
\end{table}
}

\usepackage{algorithm}
\usepackage[noend]{algpseudocode}

\newcommand{\ie}{{i.e.\ }}
\newcommand{\eg}{{e.g.\ }}
\newcommand{\citefig}[1]{Fig.~\ref{#1}}
\newcommand{\citeeq}[1]{{Eq.~\eqref{#1}}}

\newcommand{\citetable}[1]{{Table~\ref{#1}}}
\newcommand{\citesec}[1]{{Section~\ref{#1}}}

\newcommand{\argmin}[1]{\underset{#1}{\operatorname{arg}\,\operatorname{min}}\;}

\newcolumntype{L}[1]{>{\raggedright\let\newline\\\arraybackslash\hspace{0pt}}m{#1}}
\newcolumntype{C}[1]{>{\centering\let\newline\\\arraybackslash\hspace{0pt}}m{#1}}
\newcolumntype{R}[1]{>{\raggedleft\let\newline\\\arraybackslash\hspace{0pt}}m{#1}}

\usepackage{xstring}
\newcommand{\paper}[1]{{{\ignorespaces
  Paper~\textbf{\IfEqCase{#1}{%
    {pes}{[\ref{paper:pes}]}%
    {isgt}{[\ref{paper:isgt}]}%
    {freezer}{[\ref{paper:freezer}]}%
    {repl}{[\ref{paper:repl}]}%
    {upec}{[\ref{paper:upec}]}%
    {cdc}{[\ref{paper:cdc}]}%
    {gm}{[\ref{paper:gm}]}}[\PackageError{paper}{Paper not defined: #1}{}]%
}}}}

\begin{document}

\title{Computation of Ultra-Short-Term Prediction Intervals of the Power Prosumption in Active Distribution Networks}

\author{Plouton Grammatikos,
        Fabrizio~Sossan,
        Jean-Yves~Le~Boudec,
        and Mario~Paolone
\thanks{F. Sossan is at the Haute \'Ecole sp\'ecialis\'ee de Suisse occidentale Sion, P. Grammatikos, M. Paolone (Distributed Electrical Systems Laboratory) and J.Y. Le~Boudec (Laboratory for Communications and Applications) are at the \'Ecole Polytechnique F\'ed\'erale de Lausanne, Switzerland. Email: fabrizio.sossan@hevs.ch, \{plouton.grammatikos, mario.paolone, jean-yves.leboudec\}@epfl.ch.}
}

\maketitle

\begin{abstract}
Microgrids and, in general, active distribution networks require ultra-short-term prediction, i.e., for sub-second time scales, for specific control decisions. Conventional forecasting methodologies are not effective at such time scales. To address this issue, we propose a non-parametric method for computing ultra short-term prediction intervals (PIs) of the power prosumption of generic electrical-distribution networks. The method groups historical observations into clusters according to the values of influential variables. It is applied either to the original or to the differentiated power-prosumption time series. The clusters are considered statistically representative pools of future realizations of power prosumption (or its derivative). They are used to determine empirical PDFs and, by extracting the quantiles, to deliver PIs for respective arbitrary confidence levels. The models are validated a posteriori by carrying out a performance analysis that uses experimentally observed power-prosumption for different building types, thus allowing the identification of the dominant model.
\end{abstract}


%
\IEEEpeerreviewmaketitle

\section{Introduction} 
\label{sec:introduction}
After being the mainstream framework for the integration and coordination of distributed generation, the concepts of an active distribution network (ADN) and a microgrid recently came to prominence to tackle the challenges caused by the large-scale integration of variable renewable generation. 
ADNs comprise low-voltage (LV) or medium-voltage (MV) electrical grids with systems in place to control a combination of distributed energy resources (DERs), such as generators, loads, and storage devices \cite{ADN_definition}.

Due to the low level of aggregation, the ADN requirements for electrical-power prosumption\footnote{In power systems, the prosumption indicates the aggregated power provided or consumed by users that have the capability to generate electricity by means of user-owned distributed generation locally.} forecasting are different than for conventional large interconnected grids. 
One example relates to the possible violation of the ampacity rating of transformers, power converters, and lines due to sudden changes in the prosumption associated with the highly stochastic nature of prosumers\footnote{A node that can both absorb or inject power due to prosumption.}. A significant change in the prosumers' renewable-power generation, or a spike in load, can create power-flow variations that can exceed the rating of transformers. Whereas, a spike in current could cause the line relays to trip. This example is particularly relevant in the presence of photovoltaics (PV) and electrical-vehicle (EV) charging stations (CS) in the grid (e.g., \cite{9055235}). The former can exhibit power variations of even 60\% of their capacity in under a second \cite{7764064}, whereas the latter can cause significant load changes of hundreds of kW within a few seconds \cite{9494994}. The prediction of the prosumption can be integrated into various real-time (RT) (e.g., \cite{EPFL-ARTICLE-207742}) and model-predictive control (MPC) (e.g., \cite{6464620}) frameworks to ensure the safe operation of the ADNs and the optimal usage of their resources.

A further example is based on the capability of microgrids to operate autonomously. When connected to the external grid, microgrids can provide ancillary services to the upper grid layer \cite{hatziargyriou2014microgrids}. Whereas, in case of contingencies, they can operate islanded to the main grid and can enhance the resiliency of the supply to the local load. 
The islanding maneuver (i.e., the operation sequence for bringing a microgrid from connected to off-grid) requires a prediction of the prosumption in the range of the fundamental frequency period (20~ms) in order to correctly set the gains of the slack resource droop control (e.g., \cite{8442570}). 

A third example, which is of high importance both in distribution and transmission networks, is that of voltage sags \cite{9662640, DOSSANTOS2016976}. Voltage sags are defined as a sudden reduction of the voltage between 90\% and 10\% of the nominal value and can last from 10~ms up to 1~minute. They are caused primarily by power-system faults, such as short circuits, or by the start-up of large motors and can cause system outages if not treated in time. For distribution systems, where voltage sags typically last between 90-2000~ms \cite{bollen2006signal}, in order to mitigate the voltage drop, a control framework equipped with a forecasting tool acting in the sub-second range could have a timely reaction to the voltage sag by injecting an optimally computed active/reactive power into the grid.


As ultra-short-term power-prosumption forecasts are actionable for fundamental decisions in the context of ADN/microgrid operation and their RT control, we note that well-established forecasting methodologies (e.g., developed using several techniques, such as regression-based model, artificial neural network \cite{cho1995customer, Mamlook20091239, SOSSAN_FREEZER, sossan_epfl, azadeh2008annual, alfares2002electric, hahn2009electric, 8946642}) are not suited to this purpose because, besides referring to point predictions, they were developed considering a high level of aggregation and forecasting horizon from 15 minutes and up. Moreover, in certain applications, such as robust optimization (e.g., \cite{WANG2019400}), worst-case analysis is required; therefore, point predictions are inadequate.

As stated in \cite{7232358} and further supported by the experimental measurements of this paper, when decreasing the aggregation level and measurements sampling time, the power-prosumption volatility and noise level become prominent because consumer behaviors tend not to cancel out.
As the current state of the art appears to be inadequate to deliver ultra-short-term power-prosumption prediction intervals (PIs) of ADN prosumers, we propose an adaptive non-parametric method based on pattern recognition. The algorithm is designed to be computationally efficient, thus allowing for the delivery of high-time resolution probabilistic PIs in RT and at a high sampling rate with low computational overhead. The model is initially trained using a time series of the aggregated-power prosumption without requiring any knowledge of the nature and number of loads/generators present in the network. With an efficient updating and aging procedure, it is then continuously updated as new measurements become available.

The paper is organized as follows. In \citesec{sec:statement}, the problem of estimating the power prosumption PI is stated, along with a review of methods already developed in the context of power system applications. In Section \ref{sec:models}, we describe the proposed PI models whose performance is analyzed in \citesec{sec:results} by using experimental data for different building types. Finally,  in \citesec{sec:conclusions}, we summarize the findings.

\section{Problem Statement}\label{sec:statement}
As stated in the previous section, most of the existing literature on power-prosumption forecasting is concerned with point predictions, specifically the problem of estimating the expected realization of the power prosumption for a given look-ahead time. Whereas, we target the computation of PIs; in other words, we predict, with a given confidence level, the interval where the future power-prosumption realization is expected to lie. 
Denoting the PI at the target confidence level $\alpha$ as the couple $({P}^{\downarrow\alpha},{P}^{\uparrow\alpha})$ composed by the lower and upper bound of the interval, we address the problem that consists in finding the one-step-ahead PI as a function of a sequence of $n$ historical power-prosumption measurements until the time instant $i$, specifically:
\begin{align}
 \left( {P}^{\downarrow\alpha}_{i+1|i}, {P}^{\uparrow\alpha}_{i+1|i} \right) = f(P_{i}, \dots, P_{i+1-n})
\end{align}
where $i$ is the current time interval, and $f$ is a PI estimation model. 

When using parametric point predictors (such as autoregressive integrated moving average (ARIMA) models), we can determine the PIs by estimating the variance of the model residuals and computing the quantiles for the prescribed confidence level. This procedure can be performed under the hypothesis of Gaussian \emph{iid} (independent and identically distributed) model residuals. 

In cases where this hypothesis does not hold, non-parametric methods could be considered. For example, in order to determine the PIs of the power output of a wind farm, the authors of \cite{nielsen2006using} apply quantile regression to characterize the historical residuals of a state-of-the-art point-prediction model. 

The same concept, but developed using fuzzy inference instead of quantile regression, is described in \cite{pinson2006estimation}. As far as forecasting the electrical-power prosumption is concerned, non-parametric methods have been proposed lately in \cite{6706839} and \cite{Xiong2014353}. 
In the former work, an artificial neural network (ANN) with an empirically chosen number of layers was trained using historical data to provide a 30-minute-ahead PI for a given confidence level according to the values of selected data features. 
The latter work is concerned with predicting the minimum and maximum bounds of the power consumption  by applying empirical-mode decomposition and support-vector regression to an interval-valued time signal obtained from a one-hour historical sample of power consumption. Both methods target a prediction horizon that is too long for the requirements of RT ADN/microgrid operation discussed in the introduction. 

\section{Computation of Prediction Intervals}\label{sec:models}
As it will be exhaustively described in the rest of this section, the estimation methodology for PIs consists in grouping historical power-prosumption measurements into clusters according to the value of selected influential variables. At the time of delivering a PI, the values of the influential variables are determined, thus allowing for the selection of the appropriate cluster that is finally used as the empirical PDF (probability distribution function) of the next realization. The algorithm is designed to deliver a PI in rolling RT with a minimum report rate of 20ms (longer values are also analyzed). 

The algorithm operation sequence is sketched in \citefig{fig:stages}. The first phase, called batch training, consists of the off-line training of the estimation model that uses historical data. In the second phase, the one-step-ahead PI is delivered. And finally, in the third phase, the new progressively available measurements are used for the on-line model training. This methodology is applied in two flavors, specifically on the original and differentiated time-series, as discussed in the following two sections.

\begin{figure}[!ht]
\centering {
\scriptsize
\tikzstyle{block} = [rectangle, draw, fill=blue!0, 
    text width=15em, text centered, rounded corners, minimum height=3em, text width=8em]
\tikzstyle{block1} = [rectangle, draw, fill=blue!5, 
    text width=15em, text centered, minimum height=2em]
\tikzstyle{input} = [ellipse, draw, fill=red!5, text width=5em, text centered, minimum height=3em, node distance=3.5cm]
\tikzstyle{output} = [ellipse, draw, fill=green!5, text width=5em, text centered, minimum height=3em, node distance=3.5cm]
\tikzstyle{line} = [draw, -latex']
    
\begin{tikzpicture}[node distance = 1.25cm, auto]
    \node [block] (init) {$i = $ initial time slot index};
    \node [block1, below of=init, node distance=1cm] (batch) {Off-line batch training};
    
    \node [below of=init, node distance=1.5cm] (loop) {};
    \node [block1, below of=batch, node distance=1.0cm] (estimation) {On-line PI Estimation};
    
    \node [block, below of=estimation, node distance=1.0cm] (passing) {Rolling index $i=i+1$};
    \node [block1, below of=passing, node distance=1.0cm] (online) {On-line training};
ter    
    \node [input, left of=batch] (AA) {Historical data};
    \node [output, left of=estimation] (AB) {A PI is produced};
    \node [input, left of=online] (AC) {New realization};
    
    \path [line] (init) -- (batch);
    \path [line] (batch) -- (estimation);
    \path [line] (estimation) -- (passing);
    \path [line] (passing) -- (online);
    \path [line] (online) -- +(2.5,0) -- (+2.5,0 |- loop) -- (loop);
    
    \path [line] (AA) -- (batch);
    \path [line] (estimation) -- (AB);
    \path [line] (AC) -- (online);
    
\end{tikzpicture}
}
\captionsetup{font=footnotesize,labelfont=footnotesize}
\caption{Operation sequence of the PI estimation models. The batch training is performed off-line, whereas the PIs computation and on-line training are performed in rolling RT. Ellipses denote the input and output of each phase.}\label{fig:stages}
\end{figure}
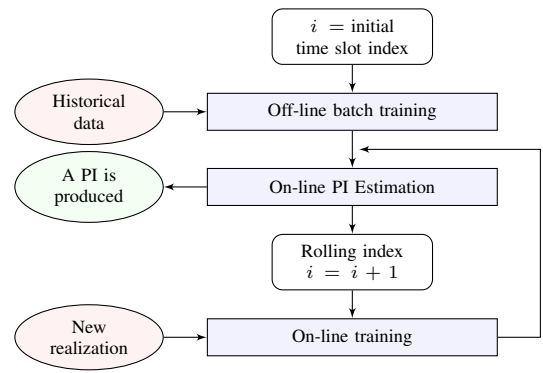

\subsection{PI Model A}

\subsubsection{Off-Line Batch Training}\label{sec:btrain}
We consider $n$ historical power-prosumption measurements $P_{i+1-n}, \dots, P_{i}$ and the respective time stamp $t_{i+1-n}, \dots, t_{i}$. Power-prosumption measurements are discretized in time and amplitude with user-defined discretization steps $\Delta T$ and $\Delta P$, respectively.
We define $c$ as the operator that determines a non-negative integer $l$, said \emph{label}, according to the power-prosumption value and timestamp in the following way:
\begin{align}
 l_j = c(P_j, t_j), \;\; l_j \in \{0,1,\dots,L-1\}\label{eq:classification}.
\end{align}
where $j$ is a generic time slot, and $L$ is the total number of labels.
The process used to operate the classification $c$ is described in detail in \citesec{sec:classification}.
The label in \eqref{eq:classification} is utilized to group the historical power-prosumption measurements $\mathcal{P}_{i,n}$ into $L$ clusters, denoted as $\mathcal{G}^0_i, \dots, \mathcal{G}^{L-1}_i$. Each cluster contains all the historical power-prosumption measurements for which the previous observation was of the respective given label. For example, the cluster $\mathcal{G}^0_i$ contains the measurements until the time slot $i$ for which the respective previous realization was with label $0$, $\mathcal{G}^1_i$ those for which the respective previous realization was with label 1, and so on. Formally, the clusters are defined as:
\begin{align}
\begin{aligned}
\mathbb{G}^l_i = \{ P_{j+1}:c(P_{j}, t_{j}) = l, \; j=i-n,\dots,i\},\\
l=0,\dots,L-1.
\end{aligned}\label{eq:clusters}
\end{align}
Let $\mathcal{G}^0_i, \dots, \mathcal{G}^{L-1}_i$ be the normalized histogram of each cluster computed as
\begin{align}
 \mathcal{G}^l_i(x) = \frac{1}{|\mathbb{G}^l_i|}\sum_{p \in \mathbb{G}^l_i} \delta(x-p), \;\; l=0,\dots, L-1\label{eq:histograms}
\end{align}
where $|\cdot|$ denotes the set cardinality (\ie, the number of elements it contains) and $\delta$ is the Dirac measure:
\begin{align}
 \delta(x) = \begin{cases}
  1 & x=0 \\
  0 & \text{otherwise}.
  \end{cases}
\end{align}
As the power prosumption is bounded, say between $P_\text{min}$ and $P_\text{max}$, histograms in \eqref{eq:histograms} are defined over a finite domain. Specifically, the domain is as:
\begin{align}
 x \in \mathbb{X} = \{P_\text{min}, P_\text{min}+\Delta P, P_\text{min}+2\Delta P, \dots,  P_\text{max}\}. \label{eq:xdomain}
\end{align}
The value of $\Delta P$ is chosen as a trade-off between accuracy and computational efficiency. Indeed, the smaller the $\Delta P$ is, the more accurate the prediction of the PI is. This aspect is made clear below. However, choosing a very small step will require more memory to store all the measurements, thus resulting in a slower computation of the PI.

\subsubsection{On-Line PI Estimation} 
At time $i$, the objective of the PI estimator is to determine the PI for the time slot $i+1$ at a given arbitrary target confidence level, said $\alpha$. The underlying idea is to assume the clusters \eqref{eq:clusters} as a statistically representative pool of possible realizations of the one-step-ahead power-prosumption realization. Therefore, the normalized histograms \eqref{eq:histograms} are assumed to be discrete PDFs and used to extract the symmetric quantiles corresponding to the $\alpha$ confidence level. Let
\begin{align}
  l_i = c(P_i, t_i) \label{eq:currentlabel}
\end{align}
be the label calculated with the information at the current time instant. The PI lower and upper bounds are determined as 
\begin{align}
 {P}^{\downarrow\alpha}_{i+1|i} = (1-\alpha)/2 \text{~~quantile~of~~} \mathcal{G}_i^{l_i},\label{eq:upperquantile}\\
 {P}^{\uparrow\alpha}_{i+1|i} = (1+\alpha)/2 \text{~~quantile~of~~}  \mathcal{G}_i^{l_i} \label{eq:lowerquantile}.
\end{align}
For the sake of clarity, the lower and upper quantiles in the expressions above are approximated by, respectively,  (also see \citefig{fig:quantilecomp}): 
\begin{align}
  \inf_{x \in \mathbb{X}} \left\{x : F_{}(x) \geq (1-\alpha)/2 \right\}\label{eq:lowerquantilecomp}
\end{align}
and
\begin{align}
  \sup_{x \in \mathbb{X}} \left\{x : F_{}(x) \leq (1-\alpha)/2 \right\},\label{eq:upperquantilecomp}
\end{align}
where $F$ denotes the discrete CDF (cumulative distribution function) of $\mathcal{G}_i^l$ calculated by computing its cumulative sum.

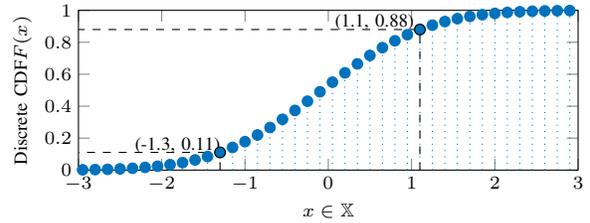
\begin{figure}[!h]
 \scriptsize 
%
%
\definecolor{mycolor1}{rgb}{0.00000,0.44700,0.74100}%
\begin{tikzpicture}

\begin{axis}[%
width=\matlabfigurewidth,
height=0.32\matlabfigurewidth,
at={(0\matlabfigurewidth,0\matlabfigurewidth)},
scale only axis,
xmin=-3,
xmax=3,
xlabel={$x \in \mathbb{X}$},
ymin=0,
ymax=1,
ylabel={$\text{Discrete CDF} F(x)$},
axis background/.style={fill=white}
]
\addplot[ycomb,color=mycolor1,dotted,mark=*,mark options={solid},forget plot] plot table[row sep=crcr] {%
-4	5.1e-05\\
-3.85	9.7e-05\\
-3.7	0.000162\\
-3.55	0.000258\\
-3.4	0.000447\\
-3.25	0.000748\\
-3.1	0.001237\\
-2.95	0.002007\\
-2.8	0.003203\\
-2.65	0.005036\\
-2.5	0.007726\\
-2.35	0.011577\\
-2.2	0.016824\\
-2.05	0.024127\\
-1.9	0.033967\\
-1.75	0.046941\\
-1.6	0.063761\\
-1.45	0.084915\\
-1.3	0.110629\\
-1.15	0.141937\\
-1	0.17813\\
-0.85	0.219784\\
-0.7	0.266515\\
-0.55	0.318033\\
-0.4	0.3735\\
-0.25	0.431087\\
-0.1	0.490521\\
0.0499999999999994	0.550297\\
0.199999999999999	0.608788\\
0.35	0.664944\\
0.5	0.717817\\
0.649999999999999	0.766403\\
0.799999999999999	0.809381\\
0.949999999999999	0.847385\\
1.1	0.87984\\
1.25	0.907301\\
1.4	0.929822\\
1.55	0.947923\\
1.7	0.96199\\
1.85	0.972757\\
2	0.980903\\
2.15	0.986871\\
2.3	0.991198\\
2.45	0.994202\\
2.6	0.99628\\
2.75	0.997648\\
2.9	0.99853\\
3.05	0.99909\\
3.2	0.999464\\
3.35	0.999717\\
3.5	0.999837\\
3.65	0.999909\\
3.8	0.999948\\
3.95	1\\
};
\addplot [color=black,only marks,mark=o,mark options={solid},forget plot]
  table[row sep=crcr]{%
-1.3	0.110629\\
};
\addplot [color=black,dashed,forget plot]
  table[row sep=crcr]{%
-1.3	0\\
-1.3	0.110629\\
-3	0.110629\\
};
\addplot [color=black,solid,mark=o,mark options={solid},forget plot]
  table[row sep=crcr]{%
1.1	0.87984\\
};
\addplot [color=black,dashed,forget plot]
  table[row sep=crcr]{%
1.1	0\\
1.1	0.87984\\
-3	0.87984\\
};
\node[right, align=left, text=black]
at (axis cs:-0,0.925) {(1.1, 0.88)};
\node[right, align=left, text=black]
at (axis cs:-2.4,0.156) {(-1.3, 0.11)};
\end{axis}
\end{tikzpicture}%
 \captionsetup{font=footnotesize,labelfont=footnotesize}
 \caption{Exemplification of the procedure \eqref{eq:lowerquantilecomp}-\eqref{eq:upperquantilecomp} to approximate the quantiles for \eqref{eq:lowerquantile}-\eqref{eq:upperquantile}. In this case, the target confidence level $\alpha$ is 80\%. }\label{fig:quantilecomp}
\end{figure}

There are two advantages to this approach. First, histograms carry the complete information over the empirical PDFs, thus allowing for computing PIs at arbitrary confidence levels by training only one model. Second, it overcomes the problem of quantile crossing that arises, for example, in \cite{nielsen2006using}, from treating the bounds at a given confidence level as two different time series.

\subsubsection{On-Line Training}
as time passes, new measurements become available and can be included to improve future PI estimates. Once the outcome $P_{i+1}$ is known, the normalized histogram associated with the label $l_i$ is updated with the new information, and the other histograms stay the same. Formally, the training procedure is as follows:
\begin{align}
 \mathcal{G}_{i+1}^{l}(x) = 
 \begin{cases}
  \hfil \phi \mathcal{G}_i^{l}(x) + (1-\phi) \delta(x-P_{i+1}) & l = l_i \\
  \hfil \mathcal{G}_i^{l}(x) & l \neq l_i ,
 \end{cases}
\end{align}
where 
\begin{align}
    \label{eq:forgetting-factor}
    \phi=\frac{T_\phi/T}{T_\phi/T+1}
\end{align} is called the \emph{forgetting factor}, $T$ is the measurement period and $T_\phi$ is called the \emph{forgetting time constant}. The forgetting factor controls how much past measurements influence the computation of PIs. Specifically, each new measurement has the same weight in the computation as all the measurements in the past $T_\phi$ seconds. The adoption of such a forgetting factor is important in order to track changes in the composition of prosumers' load/generation patterns.

\subsection{PI Model B}

\subsubsection{Off-Line Batch Training }
we apply the same principles described for Model~A but on the once differentiated power-prosumption training data-set. The differentiated time series is denoted as:
\begin{align}
\label{eq:power-differentials}
 B_{j} = P_j - P_{j-1},\;\; j=i-n+1,\dots,i.
\end{align}
The observations clusters are now calculated as follows:
\begin{align}
\begin{aligned}
\mathbb{H}^l_i = \{ B_{j+1}:c(P_{j}, t_{j}) = l, \; j=i-n+1,\dots,i\},\\
l=0,\dots,L-1,
\end{aligned}
\end{align}
and are used to determine the normalized histograms of the differentiated power prosumption time series, which are denoted as $\mathcal{H}^0_i,\dots,\mathcal{H}^{L-1}_i$.

\subsubsection{On-Line PI Estimation}
The one-step-ahead PI bounds are computed as:
\begin{align}
 {P}^{\downarrow\alpha}_{i+1|} = P_i + (1-\alpha)/2 \text{ quantile of } \mathcal{H}_i^{l_i},\\
 {P}^{\uparrow\alpha}_{i+1|i} = P_i + (1+\alpha)/2 \text{ quantile of } \mathcal{H}_i^{l_i},
\end{align}
i.e., the current power-prosumption plus two back-off terms representing the expected power prosumption variation with respect to the current realization $P_i$.

\subsubsection{On-Line Training}
Once the prosumption $P_{i+1}$ is known and the power difference $B_{i+1}$ is computed, the normalized histogram corresponding to the current label $l_i$ is updated by adding the new differenced value, whereas the others stay the same, \ie\:
\begin{align}
 \mathcal{H}_{i+1}^{l}(x) = 
 \begin{cases}
  \hfil \phi \mathcal{H}_i^{l}(x) + (1-\phi) \delta(x-B_{i+1}) & l = l_i \\
  \hfil \mathcal{H}_i^{l}(x) & l \neq l_i.
 \end{cases}
\end{align}

\subsection{Classification According to Influential Variables} \label{sec:classification}
The assignment performed through the function $c(\cdot)$ in \eqref{eq:classification} is realized by first clustering the historical measurements into groups based on similarities with respect to chosen influential variables and, then, assigning a cluster label to each new measurement. Needless to say, influential (or explanatory) variables are quantities that have an influence on the power prosumption. In general, they can be discovered using numerical methods (like analysis of variance, correlation analysis, or other procedures \cite{chernoff2009discovering}) or identified by exploiting any empirical knowledge on the observed process (in our case the structure of prosumption of a given node).

\begin{figure}[!h]
 \input{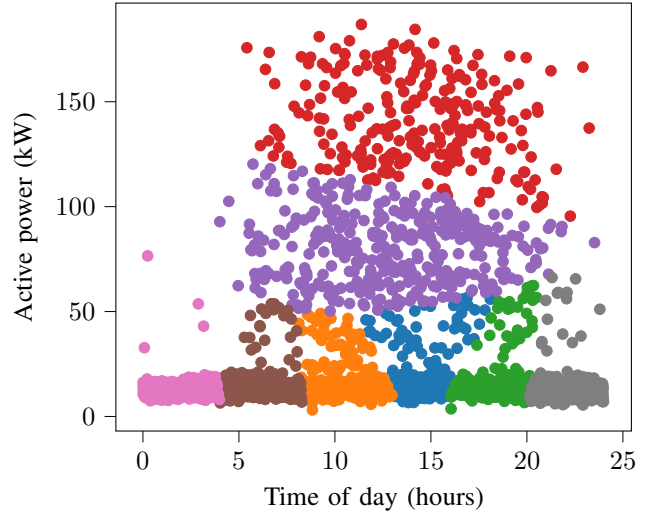}
 \captionsetup{font=footnotesize,labelfont=footnotesize}
 \caption{Visualization of k-means clustering for a node with an EV charging station with 8 clusters and using the power level and the time of day as influential variables. Positive power is consumption.}\label{fig:k-means-clustering}
\end{figure}

In this work, influential variables are chosen by adopting the latter approach. The chosen variables are (i) power magnitudes and (ii) time of day (in seconds). The first variable accounts that, in a limited capacity feeder, prosumption variations depend on the same power-prosumption magnitude. For example, when the consumption is large, a load disconnection is more likely than a load insertion because many loads are already active, and vice-versa. The second variable is supposed to capture the different power-prosumption patterns that might occur during the day.

The historical measurements are grouped, by using a k-means approach, into $L$ clusters, according to the values of the influential variables. An example of the clustering method is shown on \citefig{fig:k-means-clustering} for a node with a level-3 EV charging station. The k-means algorithm was trained with two weeks of historical measurements by using both the power level and the time of day as influential variables and by grouping the measurements into $L=8$ clusters, each of which is labeled with a different color in the figure. The value of $L$ is chosen by the user and is fixed a priori. Its influence on the performance of the algorithm is studied in Section~\ref{sec:results}.

During the off-line and on-line training phases of the algorithm, the label of measurement $(P_i, t_i)$ is computed as follows:
\begin{equation}
    l_i = \argmin {l=0,...,L-1} d(e(P_i, t_i), c_l)
    \label{eq:label-computation}
\end{equation}
where $e(P_i, t_i)$ is the point in the $m-d$ space whose coordinates are the  $m$ influential variables chosen in the classification scheme,  $c_l$ is the center of cluster $l$ and $d(\cdot, \cdot)$ is the Euclidean distance between two points.

The different classification schemes are introduced with the objective of performing an a-posteriori validation of the selection process of influential variables and the number of clusters. Indeed, as stated earlier in this section, influential variables are assigned by exploiting the empirical knowledge of the process: By comparing the performance of different classification schemes (in \citesec{sec:results}), it is possible to infer whether the progressively more complex classification schemes are meaningful or not.

\subsection{Implementation Aspects and Complexity}
The main design requirement of the proposed algorithm is to deliver PIs in RT in order to, for example, activate enough capacity in inertia-less microgrids or to assist in the decision process of setting the droop controller of slack generators in the islanding maneuver \cite{8442570}. 
Given the large PIs reporting rate, computational complexity is a central aspect and hence is addressed in this section. The batch-training phase does not have RT requirements because it is performed off-line. This consists in labeling each observation of the training data-set by applying the discussed classification algorithm. As the number of labels $L$ is fixed by design, the overall complexity of the classification procedure, given by \citeeq{eq:label-computation}, for one observation is constant time, or $O(1)$. Iterating it over a set of $N$ training data is an operation with linear time complexity, or $O(N)$.

The computation of the PIs and the on-line training are performed in rolling RT.
The former operation requires computing the label in \eqref{eq:currentlabel}, which is $O(1)$, and the PI bounds by \eqref{eq:upperquantile}-\eqref{eq:upperquantilecomp}, which involve a minimum and maximum search over the set $\mathbb{X}$, a problem with log time complexity with respect to the set cardinality, $O(\log_2|\mathbb{X}|)$, e.g., using a binary search that can exploit the monotonicity of the discrete CDFs. However, as the cardinality of $\mathbb{X}$ is fixed by design in \eqref{eq:xdomain}, the complexity of the problem can be regarded as constant time. 
Therefore, delivering PIs and performing on-line training are procedures whose complexities do not scale with the size of the problem. 

The training data and progressively incoming measurements are encoded in $L$ normalized histograms. Each of them is stored using $2 \times |\mathbb{X}|$ doubles, specifically the height and value of each bin. For example, assuming a discretization of 1024 levels (10~bit), preserving the information for a 1~year at 20~ms of resolution with the proposed method requires 128~kB per label (considering a double representation of 64-bit), while storing the individual values would require approximately 14~Gb.

\section{Performance Evaluation}\label{sec:results}
\subsection{A real case application: university buildings}
To test the performance of the proposed models, we consider three sequences of power-prosumption measurements that were recorded from different points inside the MV distribution network of the EPFL campus. 
The first one consists of an office building with a maximum consumption of 80~kW and that is equipped with a 30~kVA roof PV; the second one includes an office load with a maximum consumption of 30~kW and a 150~kW level 3 EV charging station; and the third one is a heat pump with a maximum consumption of 1.5~MW.

The measurements are with a resolution of 20~ms and are provided by a PMU-based metering infrastructure that has been deployed on the university campus (see \cite{EPFL-CONF-203775}). We consider 45 days of historical power-prosumption measurements that span the period of September-October 2022. In each case, we consider two weeks of training data; they are used to construct the clusters (as explained in \citesec{sec:classification}) and to perform the off-line training of the algorithm. Then, the proposed PI estimation models are operated for a month (with on-line training), and the estimated PIs are validated against the latter data-set, at 20~ms resolution. Each month consists of approximately 130 million data points. The evaluation is performed in a simulated environment coded in C++. The simulations are executed in a Windows Server with 128GB RAM and an Intel Xeon Gold 6130 CPU at 2.10GHz.

\subsection{Performance Metrics}
\label{sec:metrics}
We introduce the following metrics to allow for a quantitative comparison between the performance of models and classification schemes. The first is the PI normalized averaged width (PINAW), which is as follows:
 \begin{align}
  \text{PINAW} = \dfrac{1}{N} \sum_{j=1}^{N} ({P}^{\uparrow\alpha}_j - {P}^{\downarrow\alpha}_j)/P_\text{nom}.
 \end{align}

The second metric is the PI coverage probability (PICP), \ie\ the percentage of power-prosumption realization that falls inside the predicted PI. It is as follows:
\begin{align}
 \text{PICP} = \dfrac{1}{N} \sum_{j=1}^{N} b^\alpha_j
\end{align}
where
\begin{align}
b^\alpha_j = \begin{cases}
1,  & {P}^{\downarrow\alpha}_j \le P_j \le {P}^{\uparrow\alpha}_j \\
0, & \text{otherwise}.
\end{cases}
\end{align}

Because there is a trade-off between the width of the PI and the accuracy of the model, it is imperative that we define a third metric to quantify it. The metric we chose for this purpose is a modification of the coverage width-based criterion (CWC) proposed in \cite{Khosravi2013602} and is defined as follows:
\begin{align}
    \label{eq:cwc}
    \text{CWC} = \text{PINAW} \max(1, e^{-\mu \frac{\text{PICP}-\alpha}{1-\alpha}})
\end{align}
where $\mu$ is a user-defined parameter that quantifies the trade-off between PICP and PINAW. For our experiments, we chose $\mu=\frac{log(10)}{10}$. This means that a deviation of one order of magnitude in the error rate penalizes the width of the interval by $10$ times. The same result can be achieved with a linear penalty with slope $0.9$, as can be seen in \citefig{fig:cwc_plot}. The $x$-axis is the percentage deviation of the error rate from the target error rate, whereas the $y$-axis is the ratio $\frac{\text{CWC}}{\text{PINAW}}$. We observe that the exponential function penalizes less severely the error rates within one order of magnitude in comparison to the linear function. However, with our choice, higher error rates are more penalized.

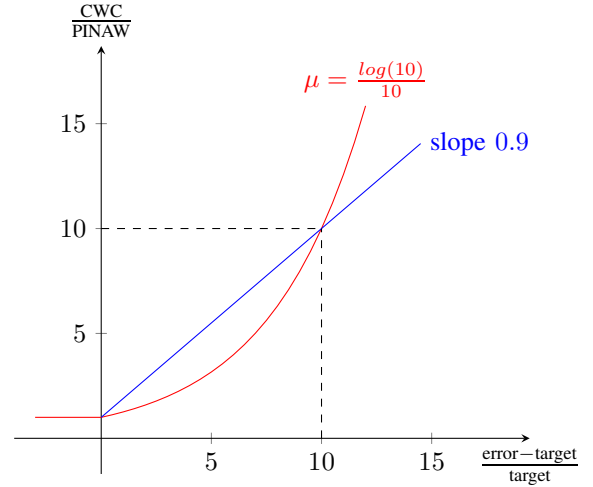
\begin{figure}
    \centering
    \begin{tikzpicture}
\begin{axis}[xmin=-2, xmax=17.5, ymin=0, ymax=17,
          axis lines=middle,
          enlargelimits,
          x label style={at={(current axis.right of origin)},anchor=north},
          y label style={at={(current axis.above origin)},anchor=south},
          xlabel={$\frac{\text{error}-\text{target}}{\text{target}}$},
          ylabel={$\frac{\text{CWC}}{\text{PINAW}}$}]
\addplot[red,domain=0:12]  {exp(ln(10)/10*x)} node[above]{$\mu=\frac{log(10)}{10}$};
\addplot[red,domain=-3:0] {1};
\addplot[blue,domain=0:14.5] {1+0.9*x} node[right]{slope $0.9$};
\draw[dashed] (0, 10) -- (10, 10);
\draw[dashed] (10, 0) -- (10, 10);
\end{axis}
\end{tikzpicture}
    \captionsetup{font=footnotesize,labelfont=footnotesize}
    \caption{Visual representation of the CWC. $\text{error}=1-\text{PICP}$ is the error rate of the algorithm, and $\text{target}=1-\alpha$ is the target error rate. The red line indicates \citeeq{eq:cwc} with the chosen value of $\mu$, and the blue line indicates a linear penalty.}
    \label{fig:cwc_plot}
\end{figure}

\subsection{Clustering Using Power Levels}
\label{sec:clustering-results}

\ResultsOfficeBuildingCWC
\ResultsChargingStationCWC
\ResultsHeatPumpCWC

To evaluate the algorithm, we consider four target confidence levels $\alpha$, namely 99, 99.9, 99.99, and 99.999\%. The parameters to be analyzed are (i) the model A or B, (ii) the number of clusters $L$ and (iii) the forgetting factor $\phi$ or, equivalently, the forgetting constant $T_\phi$. Each combination of model, $L$ and $T_\phi$ is called a \emph{configuration}. To choose the best configuration for each building, we run the algorithm for both models A and B, with all possible combinations of values of $L$ in the set $\{1, 8, 64, 256, 512, 1024\}$ and $T_\phi$ in the set $\{1, 60, 3600, 21600, 86400, 604800\}$~seconds\footnote{The values correspond to 1~second, 1~minute, 1~hour, 6~hours, 1~day and 1~week respectively.}. For all the experiments, the value of the quantization step $\Delta P$ was chosen such that the size of the domain $\mathbb{X}$ (see \citeeq{eq:xdomain}) is equal to 2000 points. It was experimentally observed that the values of PIs computed using larger domains did not change within three significant digits. Also, the average time needed to compute the PI and do one cycle of on-line training was around $50\mu s$; thus, the method meets the RT requirements.

For each combination of models, of a number of clusters, and time constants, and for each confidence level $\alpha$, we compute the three metrics PINAW, PICP, and CWC. Tables~\ref{table:elg_cwc}, \ref{table:ell_cwc}, and \ref{table:cct_cwc} show the top $5$ configurations, i.e., those that achieve the smallest CWC, for each target confidence level, for the three buildings, respectively.

Concerning the office building and the one hosting the EV charging station, model B outperforms model A for every confidence level. Also, for confidence levels up to $99.9\%$, the algorithm is able to achieve the target confidence. For higher confidence levels, however, the algorithm misses the target confidence by at most one order of magnitude. Given our choice of the parameter $\mu$, this means that the computed CI is penalized by a maximum factor of $10$ in the computation of CWC.

Regarding the optimal configuration, fewer clusters, together with larger forgetting time constants, are better choices. More specifically, the optimal number of clusters $L$ does not need to be larger than 8, whereas a value of $T_\phi$ between one day and one week is the optimal choice. By taking a closer look at Tables~\ref{table:elg_cwc} and \ref{table:ell_cwc}, we observe a trade-off between the values of $L$ and $T_\phi$ in the performance of the algorithm. If we focus on \citetable{table:elg_cwc_99}, for example, we notice that we could choose, without affecting the value of CWC by more than $1\%$, either a combination of $8$ clusters and a forgetting time constant of one week, or $256$ clusters and one day, or $1024$ clusters and $6$ hours. In fact, this table showcases that the choice of $L$ and $T_\phi$ has a minor effect on the computation of PIs for the office building. The performance of the algorithm is mainly influenced by the differentiation of the measurements performed by model B. For the charging station, however, more clusters perform up to $30\%$ worse compared to the optimal case of only $8$ clusters. 

The results are quite different for the heat pump in \citetable{table:cct_cwc}. First of all, the algorithm can only approach the target confidence levels within one order of magnitude. Unlike the other two buildings, the optimal results are achieved by a combination of few clusters (ideally only one) and a small forgetting time constant (less than an hour). This implies that the consumption of the heat pump changes more rapidly than that of the office building and the charging station, and that the measurements older than one hour do not influence the computation of PIs. We also observe that model A outperforms model B for large confidence levels. Even though the value of PINAW computed by model A is up to $5$ times larger than the one computed by model B, model B fails to achieve the target confidence level by up to two orders of magnitude, which results in an exponential increase in the metric CWC.

Overall, we conclude that the optimal choice of model, $L$ and $T_\phi$ depends on the characteristics of the prosumer. Nodes with low volatility, such as an office building, benefit from a long memory of up to one week, whereas nodes characterized by rapid power changes require the use of short memory. The clustering of power measurements seems to benefit mainly nodes with clearly distinct power-levels, such as those at a charging station. But in any case, the number of clusters does not need to be more than $8$. Finally, the differentiation of power (i.e., model B) improves the computation of PIs, provided that the power differentials have low volatility, as is the case for the office building and the charging station.

\subsection{Clustering Using Power Levels and Time of Day}

\ResultsTimeOfDay

In this section, we consider whether adding the time of day (TOD) as a feature of the algorithm, in addition to the power level (P), would improve the performance of the algorithm. The idea is that the clustering based on TOD will capture the patterns that the prosumption exhibits during the day and that this might affect the computation of the PIs. We re-run the experiments of the previous section, with the addition of the TOD. The best configuration is again the one that achieves the smallest value of CWC for each target confidence level. The comparison between the two cases, specifically (i) using only power level (P) and (ii) using power level and the time of day (P+TOD), is shown on \citetable{table:time-of-day}.

We observe that the best configuration for the two clustering strategies is the same in most cases. For the office building, the introduction of the TOD does not affect the value of CWC by more than $1\%$, which confirms once again that the clustering method does not affect the performance of the algorithm. For the charging station, the value of CWC differs by $5-10\%$ between the two clustering methods. However, the introduction of the new feature could either improve or worsen the performance, depending on the confidence level. Therefore, we cannot conclude whether it consistently performs better. As far as the heat pump is concerned, the usage of the TOD in the clustering has no effect on the performance, because the algorithm performs better when all measurements are put in one cluster.

The results indicate that this feature does not play a critical role in the computation of PIs. Perhaps other features, apart from the power level and the time of day, could influence the performance of the algorithm. The effect of additional features could be studied in future research.

\subsection{Effect of the Measurement Period}

\begin{figure*}
    \captionsetup{font=footnotesize,labelfont=footnotesize}
    \centering
    \subfloat[Office building] {
        \scriptsize 
%
%
\definecolor{mycolor1}{rgb}{0.00000,0.44700,0.74100}%
\definecolor{mycolor2}{rgb}{0.85000,0.32500,0.09800}%
\definecolor{mycolor3}{rgb}{0.92900,0.69400,0.12500}%
\definecolor{mycolor4}{rgb}{0.49400,0.18400,0.55600}%
\begin{tikzpicture}

\begin{axis}[%
width=0.6\matlabfigurewidth,
height=0.419\matlabfigurewidth,
at={(0\matlabfigurewidth,0.581\matlabfigurewidth)},
scale only axis,
xmode=log,
xmin=10,
xmax=1000000,
xminorticks=true,
xlabel style={font=\color{white!15!black}},
xlabel={Measurement period (ms)},
ymin=0,
ymax=0.6,
ylabel style={font=\color{white!15!black}},
ylabel={PINAW},
axis background/.style={fill=white},
legend style={at={(0.03,0.97)}, anchor=north west, legend cell align=left, align=left, draw=white!15!black}
]
\addplot [color=mycolor1, mark=*, mark options={solid, mycolor1}]
  table[row sep=crcr]{%
20	0.0214773041311784\\
500	0.0437306795964387\\
10000	0.0375349083901179\\
300000	0.0603368088023685\\
};
\addlegendentry{a=0.99}

\addplot [color=mycolor2, mark=*, mark options={solid, mycolor2}]
  table[row sep=crcr]{%
20	0.0305389145671871\\
500	0.0607122244030137\\
10000	0.0723873626785474\\
300000	0.0957348553750299\\
};
\addlegendentry{a=0.999}

\addplot [color=mycolor3, mark=*, mark options={solid, mycolor3}]
  table[row sep=crcr]{%
20	0.0495000471122895\\
500	0.105375965936211\\
10000	0.210907612292067\\
300000	0.146402820582813\\
};
\addlegendentry{a=0.9999}

\addplot [color=mycolor4, mark=*, mark options={solid, mycolor4}]
  table[row sep=crcr]{%
20	0.0976036007789109\\
500	0.14169102985542\\
10000	0.537208096468681\\
300000	0.147985019690318\\
};
\addlegendentry{a=0.99999}

\end{axis}

\begin{axis}[%
width=0.6\matlabfigurewidth,
height=0.419\matlabfigurewidth,
at={(0\matlabfigurewidth,0\matlabfigurewidth)},
scale only axis,
xmode=log,
xmin=10,
xmax=1000000,
xminorticks=true,
xlabel style={font=\color{white!15!black}},
xlabel={Measurement period (ms)},
ymode=log,
ymin=1e-06,
ymax=0.01,
ytick={ 1e-06,  1e-05, 0.0001,  0.001,   0.01},
yminorticks=true,
ylabel style={font=\color{white!15!black}},
ylabel={Error rate},
axis background/.style={fill=white}
]
\addplot [color=mycolor1, mark=*, mark options={solid, mycolor1}, forget plot]
  table[row sep=crcr]{%
20	0.00870778098085567\\
500	0.00995754371202062\\
10000	0.00972045464519244\\
300000	0.00779192273924501\\
};
\addplot [color=mycolor2, mark=*, mark options={solid, mycolor2}, forget plot]
  table[row sep=crcr]{%
20	0.000907147982782464\\
500	0.000942233739428255\\
10000	0.000943520428314359\\
300000	0.000658472344161498\\
};
\addplot [color=mycolor3, mark=*, mark options={solid, mycolor3}, forget plot]
  table[row sep=crcr]{%
20	0.00021862450699317\\
500	8.55744983606632e-05\\
10000	5.48558388554499e-05\\
300000	0.000109745390693639\\
};
\addplot [color=mycolor4, mark=*, mark options={solid, mycolor4}, forget plot]
  table[row sep=crcr]{%
20	1.20462534882737e-05\\
500	2.48678029424187e-05\\
10000	3.65705592364485e-06\\
300000	0.000109745390693639\\
};
\end{axis}
\end{tikzpicture}%
        \label{fig:resolution_elg}
    }
    \subfloat[Charging station] {
        \scriptsize 
%
%
\definecolor{mycolor1}{rgb}{0.00000,0.44700,0.74100}%
\definecolor{mycolor2}{rgb}{0.85000,0.32500,0.09800}%
\definecolor{mycolor3}{rgb}{0.92900,0.69400,0.12500}%
\definecolor{mycolor4}{rgb}{0.49400,0.18400,0.55600}%
\begin{tikzpicture}

\begin{axis}[%
width=0.6\matlabfigurewidth,
height=0.419\matlabfigurewidth,
at={(0\matlabfigurewidth,0.581\matlabfigurewidth)},
scale only axis,
xmode=log,
xmin=10,
xmax=1000000,
xminorticks=true,
xlabel style={font=\color{white!15!black}},
xlabel={Measurement period (ms)},
ymin=0,
ymax=0.8,
ylabel style={font=\color{white!15!black}},
ylabel={PINAW},
axis background/.style={fill=white},
legend style={at={(0.03,0.97)}, anchor=north west, legend cell align=left, align=left, draw=white!15!black}
]
\addplot [color=mycolor1, mark=*, mark options={solid, mycolor1}]
  table[row sep=crcr]{%
20	0.00706656118742768\\
500	0.0122313935495099\\
10000	0.0226826536828473\\
300000	0.208485563721533\\
};

\addplot [color=mycolor2, mark=*, mark options={solid, mycolor2}]
  table[row sep=crcr]{%
20	0.0192950937586964\\
500	0.0281412007447527\\
10000	0.125170519096175\\
300000	0.600019999412796\\
};

\addplot [color=mycolor3, mark=*, mark options={solid, mycolor3}]
  table[row sep=crcr]{%
20	0.0320194315856249\\
500	0.0561921499584666\\
10000	0.423107209529352\\
300000	0.660612779327172\\
};

\addplot [color=mycolor4, mark=*, mark options={solid, mycolor4}]
  table[row sep=crcr]{%
20	0.0518347467095\\
500	0.529763681697782\\
10000	0.573847341460918\\
300000	0.679783267267619\\
};

\end{axis}

\begin{axis}[%
width=0.6\matlabfigurewidth,
height=0.419\matlabfigurewidth,
at={(0\matlabfigurewidth,0\matlabfigurewidth)},
scale only axis,
xmode=log,
xmin=10,
xmax=1000000,
xminorticks=true,
xlabel style={font=\color{white!15!black}},
xlabel={Measurement period (ms)},
ymode=log,
ymin=1e-06,
ymax=0.0515846257585975,
ytick={ 1e-06,  1e-05, 0.0001,  0.001,   0.01},
yminorticks=true,
ylabel style={font=\color{white!15!black}},
ylabel={Error rate},
axis background/.style={fill=white}
]
\addplot [color=mycolor1, mark=*, mark options={solid, mycolor1}, forget plot]
  table[row sep=crcr]{%
20	0.00932398926222966\\
500	0.00993568526697308\\
10000	0.00980880507855653\\
300000	0.0515846257585975\\
};
\addplot [color=mycolor2, mark=*, mark options={solid, mycolor2}, forget plot]
  table[row sep=crcr]{%
20	0.00132331919836859\\
500	0.000743800027901864\\
10000	0.00335199715361134\\
300000	0.00348392897280292\\
};
\addplot [color=mycolor3, mark=*, mark options={solid, mycolor3}, forget plot]
  table[row sep=crcr]{%
20	0.000423269190014119\\
500	0.000215724479391444\\
10000	0.000108612198273406\\
300000	0.00123623286131713\\
};
\addplot [color=mycolor4, mark=*, mark options={solid, mycolor4}, forget plot]
  table[row sep=crcr]{%
20	8.29565066787774e-05\\
500	2.24712999363774e-06\\
10000	1.87262410816791e-05\\
300000	0.00101146325016854\\
};
\end{axis}
\end{tikzpicture}%
        \label{fig:resolution_ell}
    }
    \subfloat[Heat pump] {
        \scriptsize 
%
%
\definecolor{mycolor1}{rgb}{0.00000,0.44700,0.74100}%
\definecolor{mycolor2}{rgb}{0.85000,0.32500,0.09800}%
\definecolor{mycolor3}{rgb}{0.92900,0.69400,0.12500}%
\definecolor{mycolor4}{rgb}{0.49400,0.18400,0.55600}%
\begin{tikzpicture}

\begin{axis}[%
width=0.6\matlabfigurewidth,
height=0.419\matlabfigurewidth,
at={(0\matlabfigurewidth,0.581\matlabfigurewidth)},
scale only axis,
xmode=log,
xmin=10,
xmax=1000000,
xminorticks=true,
xlabel style={font=\color{white!15!black}},
xlabel={Measurement period (ms)},
ymin=0,
ymax=0.6,
ylabel style={font=\color{white!15!black}},
ylabel={PINAW},
axis background/.style={fill=white},
legend style={at={(0.03,0.97)}, anchor=north west, legend cell align=left, align=left, draw=white!15!black}
]
\addplot [color=mycolor1, mark=*, mark options={solid, mycolor1}]
  table[row sep=crcr]{%
20	0.00940420713813948\\
500	0.00569553501273139\\
10000	0.00848344882719184\\
300000	0.0658788679167607\\
};

\addplot [color=mycolor2, mark=*, mark options={solid, mycolor2}]
  table[row sep=crcr]{%
20	0.0125709633153016\\
500	0.0160608040518219\\
10000	0.0298026027562402\\
300000	0.147261008676579\\
};

\addplot [color=mycolor3, mark=*, mark options={solid, mycolor3}]
  table[row sep=crcr]{%
20	0.159491263937909\\
500	0.0260426629342155\\
10000	0.100322716713316\\
300000	0.265198753072748\\
};

\addplot [color=mycolor4, mark=*, mark options={solid, mycolor4}]
  table[row sep=crcr]{%
20	0.185312311955977\\
500	0.0736032363773233\\
10000	0.551583531059722\\
300000	0.315816104894077\\
};

\end{axis}

\begin{axis}[%
width=0.6\matlabfigurewidth,
height=0.419\matlabfigurewidth,
at={(0\matlabfigurewidth,0\matlabfigurewidth)},
scale only axis,
xmode=log,
xmin=10,
xmax=1000000,
xminorticks=true,
xlabel style={font=\color{white!15!black}},
xlabel={Measurement period (ms)},
ymode=log,
ymin=1e-05,
ymax=0.0184014450214496,
ytick={ 1e-06,  1e-05, 0.0001,  0.001,   0.01},
yminorticks=true,
ylabel style={font=\color{white!15!black}},
ylabel={Error rate},
axis background/.style={fill=white}
]
\addplot [color=mycolor1, mark=*, mark options={solid, mycolor1}, forget plot]
  table[row sep=crcr]{%
20	0.0116807522103853\\
500	0.0168110853019531\\
10000	0.0183962530331245\\
300000	0.0184014450214496\\
};
\addplot [color=mycolor2, mark=*, mark options={solid, mycolor2}, forget plot]
  table[row sep=crcr]{%
20	0.00291404936501782\\
500	0.00158228175487307\\
10000	0.00142956567537578\\
300000	0.00214495371415668\\
};
\addplot [color=mycolor3, mark=*, mark options={solid, mycolor3}, forget plot]
  table[row sep=crcr]{%
20	0.000554242342388989\\
500	0.000362277063704952\\
10000	0.000244530970787915\\
300000	0.000790246105215608\\
};
\addplot [color=mycolor4, mark=*, mark options={solid, mycolor4}, forget plot]
  table[row sep=crcr]{%
20	4.49253505974134e-05\\
500	2.61456447845365e-05\\
10000	4.13821642871737e-05\\
300000	0.000112892300745071\\
};
\end{axis}
\end{tikzpicture}%
        \label{fig:resolution_cct}
    }
    \caption{Performance evaluation as a function of the measurement period}
    \label{fig:resolution-performance}
\end{figure*}
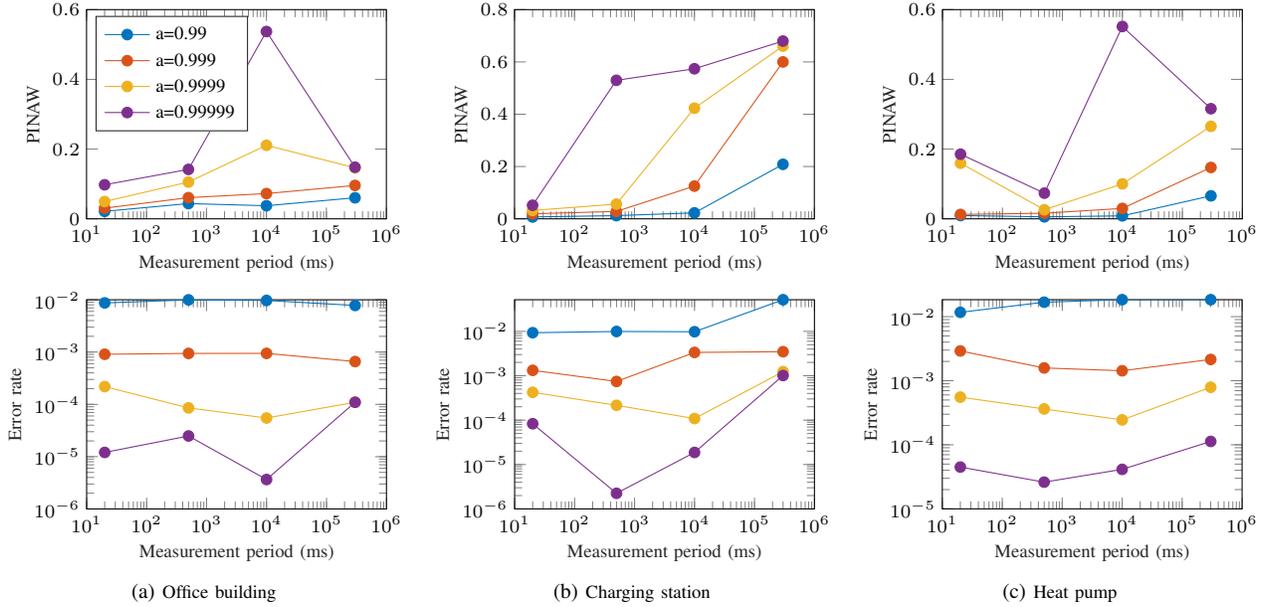

\ResultsResolution

A crucial objective of this work is to find out how the proposed algorithm scales as the measurement period increases. To test the algorithm on different measurement periods, we integrate the available data-set. In particular, given the measurements $P_j^{20}, j=1..N$ at $20ms$ resolution, the measurements at resolution $T~(ms)$ are recomputed as:
\begin{equation}
    P_i^{T} = \frac{20}{T} \sum_{j=(i-1)\frac{T}{20}+1}^{i\frac{T}{20}} P_j^{20}, i=1..\frac{20}{T}N
\end{equation}
where $T$ is assumed to be a multiple of $20ms$. The forgetting factor $\phi$ is scaled according to $T$, as in \citeeq{eq:forgetting-factor}. For this and the following section, we do clustering based only on the power level.

We perform again simulations with varying cluster numbers $L$ and forgetting time constants $T_\phi$ for different confidence levels. In \citefig{fig:resolution-performance}, we plot the values of PINAW and PICP as a function of the measurement period from $20ms$ up to $300s$ for the three buildings considered. Each point on the graphs corresponds to a different configuration (the one that achieves the smallest CWC), as shown in \citetable{table:config-resolution}.

The graph for the office building showcases that the algorithm performs well even for larger measurement periods, provided that the target confidence level is less than $99.99\%$. Indeed, for low confidence levels, the error rate is kept below the target, and the average width of the PI is less than $20\%$ of the nominal value. 
For larger confidence levels, however, the algorithm fails because either the target level is not achieved or the width of the PI is so high that it becomes useless for grid control. It is worth noting that at a resolution of five minutes, the result of the algorithm is identical for all target confidence levels above $99.99\%$. This implies that, when the measurements are sparse, we might need a larger history to meet high confidence levels.

Similar observations can be made for the charging station. The main difference is that the maximum measurement period with an acceptable performance depends on the target confidence level. With target confidence of $99\%$, the algorithm computes a low PI width for measurement periods up to $10sec$; whereas, with $99.9-99.999\%$, it has acceptable performance but only when the resolution is sub-second. For higher confidence, the algorithm cannot predict accurate PIs for periods larger than $20ms$.

Looking at the results for the heat pump on \citefig{fig:resolution-performance}, we notice a break in the trend of increasing PINAW. To understand why this happens, we look at \citetable{table:resolution-cct}. The best configuration for confidence levels $99.99-99.999\%$ changes from using model A at $20ms$ resolution to model B at higher resolutions. Model A has been shown in \citesec{sec:clustering-results} to generate, in general, larger PIs than model B. This increase in the width results in a lower error rate, which in turn might reduce the value of CWC, which is the sole metric used to compare configurations to one another. The value of CWC ultimately depends on the choice of the parameter $\mu$, which might affect the trends in the graphs of \citefig{fig:resolution-performance}.

\subsection{Confidence Level Uncertainty}

\begin{figure}
    \captionsetup{font=footnotesize,labelfont=footnotesize}
    \subfloat[Office building] {
        \scriptsize 
%
%
\definecolor{mycolor1}{rgb}{1.00000,0.00000,1.00000}%
\begin{tikzpicture}

\begin{axis}[%
width=0.92\matlabfigurewidth,
height=0.6\matlabfigurewidth,
at={(0\matlabfigurewidth,0\matlabfigurewidth)},
scale only axis,
unbounded coords=jump,
xmin=0.5,
xmax=4.5,
xtick={1,2,3,4},
xticklabels={{0.99},{0.999},{0.9999},{0.99999}},
xlabel style={font=\color{white!15!black}},
xlabel={Target confidence level},
ymode=log,
ymin=1e-06,
ymax=0.1,
yminorticks=true,
ylabel style={font=\color{white!15!black}},
ylabel={Error rate},
axis background/.style={fill=white}
]
\addplot [color=black, dashed, forget plot]
  table[row sep=crcr]{%
1	0.0119731481481481\\
1	0.0172851851851852\\
};
\addplot [color=black, dashed, forget plot]
  table[row sep=crcr]{%
2	0.00111342592592589\\
2	0.00198888888888893\\
};
\addplot [color=black, dashed, forget plot]
  table[row sep=crcr]{%
3	0.000316898148148115\\
3	0.000729629629629636\\
};
\addplot [color=black, dashed, forget plot]
  table[row sep=crcr]{%
4	1.85185185186398e-06\\
4	4.62962962965996e-06\\
};
\addplot [color=black, dashed, forget plot]
  table[row sep=crcr]{%
1	0.000118518518518518\\
1	0.00659953703703703\\
};
\addplot [color=black, dashed, forget plot]
  table[row sep=crcr]{%
2	5.27777777777905e-05\\
2	0.000478009259259254\\
};
\addplot [color=black, dashed, forget plot]
  table[row sep=crcr]{%
3	2.77777777779598e-06\\
3	3.86574074074386e-05\\
};
\addplot [color=black, dashed, forget plot]
  table[row sep=crcr]{%
4	0\\
4	0\\
};
\addplot [color=black, forget plot]
  table[row sep=crcr]{%
0.875	0.0172851851851852\\
1.125	0.0172851851851852\\
};
\addplot [color=black, forget plot]
  table[row sep=crcr]{%
1.875	0.00198888888888893\\
2.125	0.00198888888888893\\
};
\addplot [color=black, forget plot]
  table[row sep=crcr]{%
2.875	0.000729629629629636\\
3.125	0.000729629629629636\\
};
\addplot [color=black, forget plot]
  table[row sep=crcr]{%
3.875	4.62962962965996e-06\\
4.125	4.62962962965996e-06\\
};
\addplot [color=black, forget plot]
  table[row sep=crcr]{%
0.875	0.000118518518518518\\
1.125	0.000118518518518518\\
};
\addplot [color=black, forget plot]
  table[row sep=crcr]{%
1.875	5.27777777777905e-05\\
2.125	5.27777777777905e-05\\
};
\addplot [color=black, forget plot]
  table[row sep=crcr]{%
2.875	2.77777777779598e-06\\
3.125	2.77777777779598e-06\\
};
\addplot [color=black, forget plot]
  table[row sep=crcr]{%
3.875	0\\
4.125	0\\
};
\addplot [color=blue, forget plot]
  table[row sep=crcr]{%
0.75	0.00659953703703703\\
0.75	0.0119731481481481\\
1.25	0.0119731481481481\\
1.25	0.00659953703703703\\
0.75	0.00659953703703703\\
};
\addplot [color=blue, forget plot]
  table[row sep=crcr]{%
1.75	0.000478009259259254\\
1.75	0.00111342592592589\\
2.25	0.00111342592592589\\
2.25	0.000478009259259254\\
1.75	0.000478009259259254\\
};
\addplot [color=blue, forget plot]
  table[row sep=crcr]{%
2.75	3.86574074074386e-05\\
2.75	0.000316898148148115\\
3.25	0.000316898148148115\\
3.25	3.86574074074386e-05\\
2.75	3.86574074074386e-05\\
};
\addplot [color=blue, forget plot]
  table[row sep=crcr]{%
3.75	0\\
3.75	1.85185185186398e-06\\
4.25	1.85185185186398e-06\\
4.25	0\\
3.75	0\\
};
\addplot [color=red, forget plot]
  table[row sep=crcr]{%
0.75	0.00869907407407411\\
1.25	0.00869907407407411\\
};
\addplot [color=red, forget plot]
  table[row sep=crcr]{%
1.75	0.000764814814814829\\
2.25	0.000764814814814829\\
};
\addplot [color=red, forget plot]
  table[row sep=crcr]{%
2.75	0.000137037037037047\\
3.25	0.000137037037037047\\
};
\addplot [color=red, forget plot]
  table[row sep=crcr]{%
3.75	0\\
4.25	0\\
};
\addplot [color=black, only marks, mark=+, mark options={solid, draw=red}, forget plot]
  table[row sep=crcr]{%
nan	nan\\
};
\addplot [color=black, only marks, mark=+, mark options={solid, draw=red}, forget plot]
  table[row sep=crcr]{%
2	0.00227962962962958\\
2	0.00243703703703702\\
2	0.00261111111111112\\
2	0.00267407407407405\\
2	0.00275925925925924\\
2	0.00312500000000004\\
2	0.00312777777777773\\
2	0.00341574074074069\\
};
\addplot [color=black, only marks, mark=+, mark options={solid, draw=red}, forget plot]
  table[row sep=crcr]{%
3	0.000743518518518504\\
3	0.000760185185185169\\
3	0.000823148148148101\\
3	0.000874999999999959\\
3	0.000884259259259279\\
3	0.000911111111111085\\
3	0.00102222222222226\\
};
\addplot [color=black, only marks, mark=+, mark options={solid, draw=red}, forget plot]
  table[row sep=crcr]{%
4	6.48148148152394e-06\\
4	8.33333333338793e-06\\
4	1.11111111110729e-05\\
4	1.11111111110729e-05\\
4	1.11111111110729e-05\\
4	1.11111111110729e-05\\
4	1.38888888888689e-05\\
4	1.94444444444608e-05\\
4	2.12962962963248e-05\\
4	2.50000000000528e-05\\
4	2.50000000000528e-05\\
4	2.77777777777377e-05\\
4	3.42592592592617e-05\\
4	3.98148148148536e-05\\
4	4.44444444444025e-05\\
4	4.53703703703345e-05\\
4	4.62962962962665e-05\\
4	4.81481481481305e-05\\
4	4.90740740740625e-05\\
4	5.46296296296545e-05\\
4	6.48148148147953e-05\\
4	0.000112962962962926\\
4	0.000126851851851906\\
4	0.000131481481481455\\
4	0.000251851851851836\\
4	0.0002537037037037\\
};
\addplot [color=mycolor1, dotted, mark=square, mark options={solid, mycolor1}, forget plot]
  table[row sep=crcr]{%
1	0.01\\
2	0.001\\
3	9.9999999999989e-05\\
4	9.99999999995449e-06\\
};
\end{axis}
\end{tikzpicture}%
        \label{fig:boxplot_elg}
    }

    \subfloat[Charging station] {
        \scriptsize 
%
%
\definecolor{mycolor1}{rgb}{1.00000,0.00000,1.00000}%
\begin{tikzpicture}

\begin{axis}[%
width=0.92\matlabfigurewidth,
height=0.6\matlabfigurewidth,
at={(0\matlabfigurewidth,0\matlabfigurewidth)},
scale only axis,
xmin=0.5,
xmax=4.5,
xtick={1,2,3,4},
xticklabels={{0.99},{0.999},{0.9999},{0.99999}},
xlabel style={font=\color{white!15!black}},
xlabel={Target confidence level},
ymode=log,
ymin=1e-06,
ymax=0.1,
yminorticks=true,
ylabel style={font=\color{white!15!black}},
ylabel={Error rate},
axis background/.style={fill=white}
]
\addplot [color=black, dashed, forget plot]
  table[row sep=crcr]{%
1	0.0104083333333333\\
1	0.0126509259259259\\
};
\addplot [color=black, dashed, forget plot]
  table[row sep=crcr]{%
2	0.00175925925925929\\
2	0.00345925925925927\\
};
\addplot [color=black, dashed, forget plot]
  table[row sep=crcr]{%
3	0.000569444444444456\\
3	0.0012981481481481\\
};
\addplot [color=black, dashed, forget plot]
  table[row sep=crcr]{%
4	0.00012175925925928\\
4	0.000280555555555506\\
};
\addplot [color=black, dashed, forget plot]
  table[row sep=crcr]{%
1	0.00495370370370374\\
1	0.00773148148148151\\
};
\addplot [color=black, dashed, forget plot]
  table[row sep=crcr]{%
2	0\\
2	0.000611574074074084\\
};
\addplot [color=black, dashed, forget plot]
  table[row sep=crcr]{%
3	0\\
3	7.12962962963193e-05\\
};
\addplot [color=black, dashed, forget plot]
  table[row sep=crcr]{%
4	0\\
4	2.77777777779598e-06\\
};
\addplot [color=black, forget plot]
  table[row sep=crcr]{%
0.875	0.0126509259259259\\
1.125	0.0126509259259259\\
};
\addplot [color=black, forget plot]
  table[row sep=crcr]{%
1.875	0.00345925925925927\\
2.125	0.00345925925925927\\
};
\addplot [color=black, forget plot]
  table[row sep=crcr]{%
2.875	0.0012981481481481\\
3.125	0.0012981481481481\\
};
\addplot [color=black, forget plot]
  table[row sep=crcr]{%
3.875	0.000280555555555506\\
4.125	0.000280555555555506\\
};
\addplot [color=black, forget plot]
  table[row sep=crcr]{%
0.875	0.00495370370370374\\
1.125	0.00495370370370374\\
};
\addplot [color=black, forget plot]
  table[row sep=crcr]{%
1.875	0\\
2.125	0\\
};
\addplot [color=black, forget plot]
  table[row sep=crcr]{%
2.875	0\\
3.125	0\\
};
\addplot [color=black, forget plot]
  table[row sep=crcr]{%
3.875	0\\
4.125	0\\
};
\addplot [color=blue, forget plot]
  table[row sep=crcr]{%
0.75	0.00773148148148151\\
0.75	0.0104083333333333\\
1.25	0.0104083333333333\\
1.25	0.00773148148148151\\
0.75	0.00773148148148151\\
};
\addplot [color=blue, forget plot]
  table[row sep=crcr]{%
1.75	0.000611574074074084\\
1.75	0.00175925925925929\\
2.25	0.00175925925925929\\
2.25	0.000611574074074084\\
1.75	0.000611574074074084\\
};
\addplot [color=blue, forget plot]
  table[row sep=crcr]{%
2.75	7.12962962963193e-05\\
2.75	0.000569444444444456\\
3.25	0.000569444444444456\\
3.25	7.12962962963193e-05\\
2.75	7.12962962963193e-05\\
};
\addplot [color=blue, forget plot]
  table[row sep=crcr]{%
3.75	2.77777777779598e-06\\
3.75	0.00012175925925928\\
4.25	0.00012175925925928\\
4.25	2.77777777779598e-06\\
3.75	2.77777777779598e-06\\
};
\addplot [color=red, forget plot]
  table[row sep=crcr]{%
0.75	0.00890879629629632\\
1.25	0.00890879629629632\\
};
\addplot [color=red, forget plot]
  table[row sep=crcr]{%
1.75	0.00115324074074075\\
2.25	0.00115324074074075\\
};
\addplot [color=red, forget plot]
  table[row sep=crcr]{%
2.75	0.000275462962962936\\
3.25	0.000275462962962936\\
};
\addplot [color=red, forget plot]
  table[row sep=crcr]{%
3.75	2.08333333333588e-05\\
4.25	2.08333333333588e-05\\
};
\addplot [color=black, only marks, mark=+, mark options={solid, draw=red}, forget plot]
  table[row sep=crcr]{%
1	0.0162537037037037\\
1	0.0199888888888888\\
1	0.0204166666666666\\
1	0.0239111111111111\\
};
\addplot [color=black, only marks, mark=+, mark options={solid, draw=red}, forget plot]
  table[row sep=crcr]{%
2	0.00359166666666666\\
2	0.00364444444444445\\
2	0.00428796296296297\\
2	0.00505648148148152\\
2	0.00562870370370372\\
};
\addplot [color=black, only marks, mark=+, mark options={solid, draw=red}, forget plot]
  table[row sep=crcr]{%
3	0.00133888888888889\\
3	0.00136666666666663\\
3	0.00141666666666662\\
3	0.00141944444444442\\
3	0.00144074074074074\\
3	0.00154537037037039\\
3	0.00166388888888891\\
3	0.002112037037037\\
3	0.00232777777777782\\
};
\addplot [color=black, only marks, mark=+, mark options={solid, draw=red}, forget plot]
  table[row sep=crcr]{%
4	0.000318518518518496\\
4	0.00036851851851849\\
4	0.00041111111111114\\
4	0.000414814814814868\\
4	0.000523148148148134\\
4	0.000562037037037055\\
4	0.000572222222222196\\
4	0.000659259259259248\\
};
\addplot [color=mycolor1, dotted, mark=square, mark options={solid, mycolor1}, forget plot]
  table[row sep=crcr]{%
1	0.01\\
2	0.001\\
3	9.9999999999989e-05\\
4	9.99999999995449e-06\\
};
\end{axis}
\end{tikzpicture}%
        \label{fig:boxplot_ell}
    }
    
    \subfloat[Heat pump] {
        \scriptsize 
%
%
\definecolor{mycolor1}{rgb}{1.00000,0.00000,1.00000}%
\begin{tikzpicture}

\begin{axis}[%
width=0.92\matlabfigurewidth,
height=0.6\matlabfigurewidth,
at={(0\matlabfigurewidth,0\matlabfigurewidth)},
scale only axis,
xmin=0.5,
xmax=4.5,
xtick={1,2,3,4},
xticklabels={{0.99},{0.999},{0.9999},{0.99999}},
xlabel style={font=\color{white!15!black}},
xlabel={Target confidence level},
ymode=log,
ymin=1e-06,
ymax=0.1,
yminorticks=true,
ylabel style={font=\color{white!15!black}},
ylabel={Error rate},
axis background/.style={fill=white}
]
\addplot [color=black, dashed, forget plot]
  table[row sep=crcr]{%
1	0.0123560185185185\\
1	0.0138\\
};
\addplot [color=black, dashed, forget plot]
  table[row sep=crcr]{%
2	0.00438101851851852\\
2	0.00696574074074074\\
};
\addplot [color=black, dashed, forget plot]
  table[row sep=crcr]{%
3	0.000874537037037049\\
3	0.00197222222222226\\
};
\addplot [color=black, dashed, forget plot]
  table[row sep=crcr]{%
4	6.57407407407273e-05\\
4	0.000153703703703711\\
};
\addplot [color=black, dashed, forget plot]
  table[row sep=crcr]{%
1	0.0102157407407407\\
1	0.011211574074074\\
};
\addplot [color=black, dashed, forget plot]
  table[row sep=crcr]{%
2	0.00121388888888885\\
2	0.00237129629629629\\
};
\addplot [color=black, dashed, forget plot]
  table[row sep=crcr]{%
3	0\\
3	0.00011481481481479\\
};
\addplot [color=black, dashed, forget plot]
  table[row sep=crcr]{%
4	0\\
4	3.70370370372797e-06\\
};
\addplot [color=black, forget plot]
  table[row sep=crcr]{%
0.875	0.0138\\
1.125	0.0138\\
};
\addplot [color=black, forget plot]
  table[row sep=crcr]{%
1.875	0.00696574074074074\\
2.125	0.00696574074074074\\
};
\addplot [color=black, forget plot]
  table[row sep=crcr]{%
2.875	0.00197222222222226\\
3.125	0.00197222222222226\\
};
\addplot [color=black, forget plot]
  table[row sep=crcr]{%
3.875	0.000153703703703711\\
4.125	0.000153703703703711\\
};
\addplot [color=black, forget plot]
  table[row sep=crcr]{%
0.875	0.0102157407407407\\
1.125	0.0102157407407407\\
};
\addplot [color=black, forget plot]
  table[row sep=crcr]{%
1.875	0.00121388888888885\\
2.125	0.00121388888888885\\
};
\addplot [color=black, forget plot]
  table[row sep=crcr]{%
2.875	0\\
3.125	0\\
};
\addplot [color=black, forget plot]
  table[row sep=crcr]{%
3.875	0\\
4.125	0\\
};
\addplot [color=blue, forget plot]
  table[row sep=crcr]{%
0.75	0.011211574074074\\
0.75	0.0123560185185185\\
1.25	0.0123560185185185\\
1.25	0.011211574074074\\
0.75	0.011211574074074\\
};
\addplot [color=blue, forget plot]
  table[row sep=crcr]{%
1.75	0.00237129629629629\\
1.75	0.00438101851851852\\
2.25	0.00438101851851852\\
2.25	0.00237129629629629\\
1.75	0.00237129629629629\\
};
\addplot [color=blue, forget plot]
  table[row sep=crcr]{%
2.75	0.00011481481481479\\
2.75	0.000874537037037049\\
3.25	0.000874537037037049\\
3.25	0.00011481481481479\\
2.75	0.00011481481481479\\
};
\addplot [color=blue, forget plot]
  table[row sep=crcr]{%
3.75	3.70370370372797e-06\\
3.75	6.57407407407273e-05\\
4.25	6.57407407407273e-05\\
4.25	3.70370370372797e-06\\
3.75	3.70370370372797e-06\\
};
\addplot [color=red, forget plot]
  table[row sep=crcr]{%
0.75	0.0118083333333333\\
1.25	0.0118083333333333\\
};
\addplot [color=red, forget plot]
  table[row sep=crcr]{%
1.75	0.00325277777777777\\
2.25	0.00325277777777777\\
};
\addplot [color=red, forget plot]
  table[row sep=crcr]{%
2.75	0.000390277777777781\\
3.25	0.000390277777777781\\
};
\addplot [color=red, forget plot]
  table[row sep=crcr]{%
3.75	2.77777777777377e-05\\
4.25	2.77777777777377e-05\\
};
\addplot [color=black, only marks, mark=+, mark options={solid, draw=red}, forget plot]
  table[row sep=crcr]{%
1	0.0140731481481482\\
1	0.0144824074074074\\
1	0.0192083333333334\\
1	0.0286305555555556\\
};
\addplot [color=black, only marks, mark=+, mark options={solid, draw=red}, forget plot]
  table[row sep=crcr]{%
2	0.00793703703703708\\
};
\addplot [color=black, only marks, mark=+, mark options={solid, draw=red}, forget plot]
  table[row sep=crcr]{%
3	0.00206851851851853\\
3	0.00208055555555553\\
3	0.0022064814814815\\
3	0.00226018518518523\\
};
\addplot [color=black, only marks, mark=+, mark options={solid, draw=red}, forget plot]
  table[row sep=crcr]{%
4	0.00016296296296292\\
4	0.000168518518518512\\
4	0.000199074074074046\\
4	0.00020092592592591\\
4	0.000262962962962909\\
4	0.000277777777777821\\
};
\addplot [color=mycolor1, dotted, mark=square, mark options={solid, mycolor1}, forget plot]
  table[row sep=crcr]{%
1	0.01\\
2	0.001\\
3	9.9999999999989e-05\\
4	9.99999999995449e-06\\
};
\end{axis}
\end{tikzpicture}%
        \label{fig:boxplot_cct}
    }
    \caption{Box plot of model accuracy versus target confidence level at $20ms$ resolution}
    \label{fig:box_plots}
\end{figure}
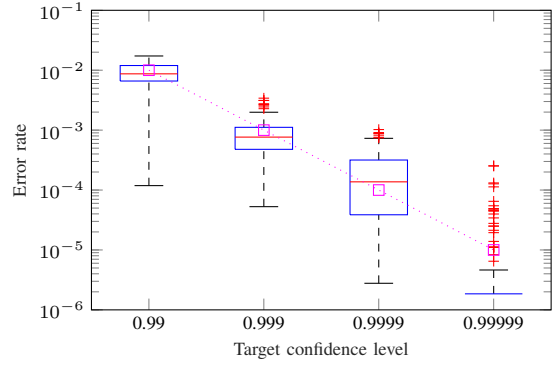
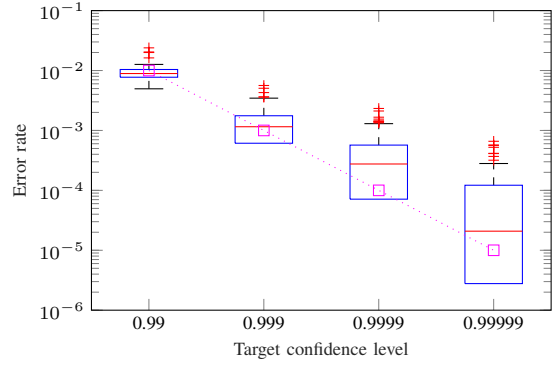
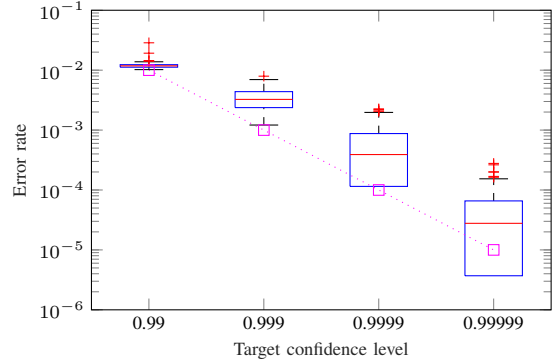

The metrics of \citesec{sec:metrics} evaluate, over the full one-month period, the average performance of the algorithm. However, it would be interesting to see how the algorithm performs over time. For this purpose, we split the one-month evaluation period into six-hour windows and compute the error rate achieved by the algorithm in each window. Hence, for each confidence level, we compute a histogram of $120$ estimations of the error rate. We then depict the statistical measures of the histograms using the box plots of \citefig{fig:box_plots}. The blue box indicates the 25th and 75th percentiles, the red line is the median, the red crosses are outliers, and the small square indicates the target error rate. We computed the box plots only for the best configuration for each building and for confidence level at $20ms$ resolution, which are shown in \citetable{table:config-resolution}.

If the algorithm is consistent in predicting accurate PIs, then the median of the box plot should be close to the target error-rate. We see that this is indeed the case for the office building. For the largest confidence level, our algorithm performs even better than expected, albeit with many outliers. For the charging station, the performance is sometimes worse than expected, but the target error-rate is nevertheless contained within the 25th and the 75th percentile of the box plot. This is not always the case for the heat pump, hence the algorithm cannot consistently estimate accurate PIs if there is high volatility in the measurements.

The results showcase that our algorithm computes accurate
PIs, irrespective of the building, provided that the target confidence level is less than $99.9\%$. For higher confidence levels, the performance is still acceptable, but the consistency of the predictions depends on the dynamics of the node.

\section{Conclusion}\label{sec:conclusions}
Motivated by the requirements of ADNs real-time control, we have presented a non-parametric method for computing ultra-short-term PIs (prediction intervals) of the power prosumption in generic ADNs (\eg buildings).
The method consists in grouping historical measurements into clusters, according to the value of selected influential variables. The clusters are considered statistically representative pools of future power-prosumption realizations and are used to extract PIs at arbitrary confidence levels by calculating the quantiles from the respective PDF. 

The proposed method has been applied to the original and once-differentiated power-prosumption time series, and different influential variables have been considered. The performance of the method was tested for different types of prosumers by using experimental measurements from an MV distribution network.
The performance analysis enabled us to make an a-posteriori selection of the parameters of the algorithm. The algorithm was shown to compute relatively narrow PIs for the studied prosumers, for time resolutions from $20ms$ up to a few minutes in some cases, provided that the target confidence level is below $99.9\%$. 

A final statement concerns the computational complexity, which becomes a relevant concern especially when considering densely sampled time series and the high reporting rate for the predictions. 
We have shown that the proposed algorithm performs the PI computation and on-line training in constant time hence is scalable.


\ifCLASSOPTIONcaptionsoff
  \newpage
\fi



\bibliographystyle{IEEEtran}
\bibliography{biblio.bib}

\end{document}